\documentstyle[psfig,aps,prl,multicol,eqsecnum]{revtex}
\begin{document}
\title{Information-theoretic interpretation of quantum error-correcting codes}

\author{Nicolas J. Cerf$^{1,3}$ and Richard Cleve$^{2,3}$}
\address{$^1$W. K. Kellogg Radiation Laboratory,
California Institute of Technology, Pasadena, California 91125\\
$^2$Department of Computer Science, University of Calgary, Calgary, Alberta,
Canada T2N 1N4\\
$^3$Institute for Theoretical Physics, University of California,
Santa Barbara, California 93106}

\date{February 1997}

\draft
\maketitle
\begin{abstract}

Quantum error-correcting codes are analyzed from an information-theoretic
perspective centered on quantum conditional and mutual entropies. This
approach parallels the description of classical error correction in Shannon
theory, while clarifying the differences between classical and quantum
codes. More specifically, it is shown how quantum information theory
accounts for the fact that ``redundant'' information
can be distributed over quantum bits even though this does not violate
the quantum ``no-cloning'' theorem. 
Such a remarkable feature, which has no counterpart for classical codes,
is related to the property that the ternary mutual entropy vanishes
for a tripartite system in a pure state. This information-theoretic
description of quantum coding is used to derive the quantum analogue of
the Singleton bound on the number of logical bits
that can be preserved by a code of fixed length which
can recover a given number of errors.

\end{abstract}
\pacs{PACS numbers: 03.65.Bz,89.70.+c
      \hfill KRL preprint MAP-209}

\begin{multicols}{2}[]
\narrowtext

\section{Introduction}

The potential use of quantum computers for solving certain classes of problems
has recently received a considerable amount of attention 
(for a review see, {\it e.g.}, \cite{bib_lloyd,bib_divinc,bib_ekertjoz}).
A major obstacle
in the building of quantum computers, however, is the coupling of the computer
with its environment or the {\it decoherence}, which rapidly destroys
the quantum superposition at the heart of quantum algorithms.
An essential element in the realization of such quantum computers
is therefore the use of quantum error-correcting codes,
which have been shown to ensure protection against 
decoherence~\cite{bib_shor,bib_steane,bib_calderbank,bib_ekert,bib_laflamme,bib_bdsw,bib_vaidman,bib_knilaf,bib_calderbank2,bib_grassl,bib_peres}.
Quantum codes are similar in many respects to classical codes.
In classical coding theory, logical words (of $k$ bits)
are encoded into codewords (of $n>k$ bits). The latter are suitably
chosen among the set of all $2^n$ possible words of $n$ bits so that
the alteration because of noise of say $t$ bits (at most) can be recovered. 
A specific set of codewords then constitutes 
an $[n,k,t]$ code, encoding $k$ bits
into $n$ bits and correcting all patterns of $t$ (or fewer) errors
among those $n$ bits.
The simplest example of a classical code with $k=1$, $n=3$, and $t=1$ is
the repetition code where a logical bit 0 (or 1) is encoded
into 000 (or 111); decoding is simply
performed using the majority rule, which is enough to recover $t=1$ errors.
In classical coding theory, corrupted data
is thus restored by introducing redundancy
($n>k$), that is by duplicating part of the information that must be
preserved. (In the above---very inefficient---example, information
is triplicated.) In quantum coding theory, the central issue
is to find a set of $2^k$ quantum codewords (of $n$ qubits) such that
{\em quantum information} can be protected against the alteration
due to coupling with an environment ({\it i.e.}, such that the quantum system 
survives decoherence).  At first sight, it seems that, since
the duplication of an arbitrary quantum state is forbidden 
by the quantum {\em no-cloning} theorem\cite{bib_noncloning},
``quantum redundancy'' is
impossible. However, after the pioneering work of Shor\cite{bib_shor},
it has been realized that quantum coding is achievable in spite of the
no-cloning theorem, and a great deal of work has recently been
devoted to this issue~\cite{bib_steane,bib_calderbank,bib_ekert,bib_laflamme,bib_bdsw,bib_vaidman,bib_knilaf,bib_calderbank2,bib_grassl,bib_peres}.
It has been shown that quantum information
can be distributed over many qubits through a suitable encoding and
subsequently recovered after partial alteration, without
violating the no-cloning theorem.
\par

In this paper, we aim at clarifying some aspects of
quantum coding from a perspective centered on quantum entropies.
It has been shown recently that classical and quantum entropies
can be described within a unified information-theoretical framework
involving negative conditional entropies~\cite{bib_neginfo,bib_measure},
as briefly outlined in Appendix~\ref{sect_app}.
Here, we apply this framework to quantum error-correcting codes,
paralleling the classical description of error correction in Shannon theory.
We show that, for an arbitrary entanglement between the logical words
and a ``reference'' system to be preserved,
the quantum {\it mutual} entropy between this ``reference''
and any ``interacting'' part of the codewords
must be vanishing  prior to decoherence. In other words,
an entropic condition for perfect quantum error correction
is that the ``reference'' system is statistically {\it independent} of
any arbitrarily chosen part of the codewords that might interact
with the environment. This condition relies on the conservation
of quantum mutual entropies implied by unitarity, along with the 
property of strong subadditivity of quantum entropies
(see Appendix~\ref{sect_app}).
It expresses the fact that the environment cannot become 
{\it directly} entangled with the ``reference'' system
(entanglement may arise only via the
codewords), or, roughly speaking, that the environment cannot extract
information about the logical words. 
\par

We continue by deriving the analogue of the Singleton bound 
for quantum codes~\cite{bib_knilaf}, {\it i.e.}, $k \le n-4t$,
using simple arguments based on this entropic approach.
Such an information-theoretic description of coding sheds new light 
on the interpretation of this bound in terms of ``weak'' cloning.
While the quantum bits that are altered
as a result of any error are statistically independent
of the reference (or the encoded logical word),
the quantum information stored in the entire codeword remains unaffected.
This results from the fact that the {\it ternary}
mutual entropy vanishes for any entangled tripartite system 
in a pure state,
a property which has no classical counterpart~\cite{bib_measure}.
The central point is that,
in contrast with classical codes, no duplicating---or full cloning---is
achieved by {\it quantum} error-correcting codes. Rather, a ``weak'' quantum
cloning is achieved, such that any part of the codeword susceptible to decohere
appears independent of the reference although the entire codeword remains
entangled with it. This purely quantum situation 
is forbidden in classical information theory due to the non-negativity
of Shannon conditional entropies, and reflects
a fundamental difference between classical and quantum
error correcting codes.

\section{Quantum error correction}

Let us consider a set of orthogonal logical states $|i_L\rangle$ 
(with $i=1,\cdots 2^k$) which are encoded into
orthogonal codewords $|i_Q\rangle$ consisting of $n$ qubits.
(The index $Q$ refers to the quantum channel
on which the codewords are sent.) 
The states $|i_L\rangle$ belong to the logical Hilbert
space ${\cal H}_L$ of dimension $d_L=2^k$ spanned by the
$k$ logical qubits, while the states
$|i_Q\rangle$ belong to ${\cal H}_Q$ of dimension
$d_Q=2^n$. We have clearly $d_Q > d_L$, which is the quantum equivalent
of classical ``redundancy'': the logical
states are encoded in some $2^k$-dimensional subspace of the
full $2^n$-dimensional Hilbert space so that part of the information
in the $n$ qubits is ``redundant''. Qualitatively speaking,
$n-k$ qubits of the codewords represent redundant information
(they are equivalent to the ``check bits'' of classical
codes~\cite{bib_ash}). In Section~IV, we will make 
this concept of quantum ``redundancy'' more quantitative.
\par

\begin{figure}[t]
\caption{Schematic model of a noisy quantum channel preceded by encoding
and followed by decoding. The logical states
(system $L$ of $k$ qubits) are entangled with the reference system $R$.
Encoding, using an ancilla $A$ of $n-k$ ``check'' qubits
initially in a $|0\rangle$
state, yields the codewords (system $Q$ of $n$ qubits). Then, $e$ qubits
($Q_e$) are ``erased'' by interacting with the environment $E$
via $U_{QE}$, while the $n-e$ remaining ones ($Q_u$) are unchanged.
Decoding, involving the ``erased'' qubits $Q_e'$  along with the
unchanged ones $Q_u$ yields the $k$ logical bits $L$ in the initial
entangled state $\psi_{RL}$ with the reference $R$. The primes refer
to the systems after environment-induced decoherence. }
\vskip 0.25cm
\centerline{\psfig{figure=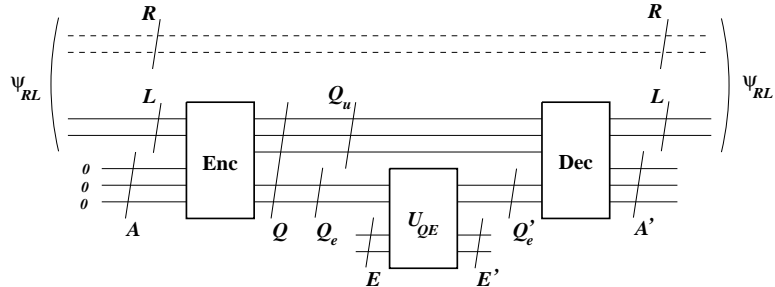,width=4.00in,angle=-90}}
\label{fig_schumacher}
\vskip 0.25cm
\end{figure}

The key property of a quantum code lies in its ability to protect
an {\em arbitrary} superposition of logical states $\sum a_i |i_L\rangle$
against decoherence. Equivalently, a quantum code is such that
the entanglement of the $k$ logical qubits with a ``reference'' 
system $R$ is preserved against decoherence. In fact,
this description of quantum coding as a mean to transmit
(or conserve) entanglement with respect to $R$ in spite of
the interaction with an environment is more convenient
for our information-theoretic description and will be adopted in the
following. Accordingly, we start by considering the
initial entangled state
\begin{equation}
|\psi_{RL}\rangle = \sum_{i=1}^{2^k} a_i |i_R\rangle |i_L\rangle 
\end{equation}
where $R$ and $L$ refer to the reference and logical states, respectively.
(This is the Schmidt decomposition of a pure entangled state.)
We then consider the transformation of $|\psi_{RL}\rangle$
due to encoding followed by decoherence.
Encoding is performed by use of a unitary transformation that maps
the states $|i_L\rangle |0\rangle$ to the codewords $|i_Q\rangle$,
where $|0\rangle$ stands for the initial state of
the $n-k$ auxiliary qubits (or check bits).
Thus, after encoding, the joint state of the reference $R$ and the
quantum channel $Q$ is
\begin{equation}  \label{eq_wavefunc}
|\psi_{RQ}\rangle = \sum_{i=1}^{2^k} a_i |i_R\rangle |i_Q\rangle 
\end{equation}
It is a pure state of vanishing entropy $S(RQ)=0$;
the quantum entropies of $R$ and $Q$ are $S(R)=S(Q)=H[a_i]$, where
$H$ stands for the Shannon entropy,
\begin{equation}
H[a_i]=-\sum_i |a_i|^2 \log |a_i|^2 \;.
\end{equation} 
Let us suppose now that the codewords are sent on a noisy quantum channel 
in which they suffer decoherence due to an environment $E$.
Following Schumacher's model of a noisy channel~\cite{bib_schum},
we assume that the environment is initially in the pure state $|0\rangle$ and
then interacts with the channel according to the unitary transformation
$U_{QE}$, so that the joint state of the entire system becomes
\begin{equation}  \label{eq_Uenvir}
|\psi_{R'Q'E'}\rangle = (1_R \otimes U_{QE}) 
\sum_{i=1}^{2^k} a_i |i_R\rangle |i_Q\rangle |0\rangle
\end{equation}
(The prime refers to the systems {\it after} decoherence.)
This noisy channel is pictured in Fig.~\ref{fig_schumacher} and
will be the basis of our description of quantum coding
in terms of quantum entropies.
More specifically, we will consider a ``deterministic'' error model
in which the position of the erroneous bits is {\em known}, usually referred
to as the quantum {\em erasure} channel~\cite{bib_grassl}.
In this channel, the decoherence induced by the environment
involves $e$ qubits at known locations, {\it i.e.}, $e$ erasures.
The component $Q_e$ (of $e$ qubits) of the codeword 
interacts with $E$ (suffers $e$ erasures), while the rest 
$Q_u$ (of $n-e$ qubits) is left unchanged by this interaction. 
Accordingly, the unitary transformation in Eq.~(\ref{eq_Uenvir})
is of the form
\begin{equation}
U_{QE}= 1_{Q_u} \otimes U_{Q_e E}
\end{equation}
As an example, we can suppose that the environment is made of $e$ qubits
initially in a $|0\rangle$ state and that $U_{Q_e E}$ effects the
exchange between these qubits and the $e$ qubits of $Q_e$ (a reversible
operation). As a result, the qubits of $Q_e$ are erased (reset to $|0\rangle$)
while the qubits of $E$ get the original value of the erased qubits.
As the environment is traced over in order to determine the state
of the channel $Q$ after decoherence, quantum information is apparently erased
even though the overall process is unitary.
Of course, any other $U_{Q_e E}$ could result from
decoherence, and a quantum erasure-correcting code will be such
that the entanglement with $R$ is preserved for an arbitrary $U_{Q_e E}$.
\par

\end{multicols}
\widetext

\begin{multicols}{2}[]
\narrowtext

Before discussing coding and decoherence using quantum entropies
(Section IV), let us first review some basics
of quantum error-correcting codes. It is known that,
rather than coupling the codewords with an environment, 
one can model the errors by use of error operators $E$. For the
purpose of error correction, it is enough to consider errors of
the type $\sigma_x$ (bit-flip), $\sigma_z$ (phase-flip), 
and $\sigma_y$ (bit- and phase-flip), since, by linearity, a code
that can correct these errors can correct arbitrary errors~\cite{bib_ekert}.
For a $[n,k,t]$ code, {\it i.e.}, a code correcting $t$ errors at most,
the error operators $E$ applied on the codewords are of the form
$1^{\otimes (n-t)}\otimes E^{\otimes t}$, {\it i.e.}, the tensor
product of the identity on $n-t$ qubits and $t$ one-bit error operators
on the altered qubits. The one-bit error operators are
any linear combinations of the algebra basis 
$\{ 1,\sigma_x,\sigma_y,\sigma_z \}$.
It has been shown by Knill and Laflamme~\cite{bib_knilaf}
that a necessary and sufficient condition on quantum error-correcting
codes is that
\begin{eqnarray}
\label{eq_KL1}
\langle i_Q | E_a^{\dagger} E_b | i_Q \rangle &=&  
\langle j_Q | E_a^{\dagger} E_b | j_Q \rangle  \\
\label{eq_KL2}
\langle i_Q | E_a^{\dagger} E_b | j_Q \rangle &=& 0
\qquad\qquad {\rm for~}i\ne j
\end{eqnarray}
where the $|i_Q\rangle$ and $|j_Q\rangle$ are any two codewords
and $E_a$, $E_b$ are chosen from the set of $t$-error operators
defined above. 
Conditions (\ref{eq_KL1}) and (\ref{eq_KL2}) can be understood
by considering the decoding operation as an ``inverse'' unitary 
transformation~\cite{bib_peres}
that maps the $n$ qubits of the corrupted codeword $Q'$
into $k$ qubits (the original logical word $L$) 
and $n-k$ check qubits (the ancilla $A'$),
as represented in Fig.~\ref{fig_schumacher}.
Considering the action of decoding on two codewords 
$|i_Q \rangle$ and $|j_Q \rangle$ that have been corrupted by
errors $E_a$ or $E_b$,
it can be shown that the state in which the ancilla is left cannot depend
on the logical state, that is the decoding must be such that
\begin{eqnarray}
E_a |i_Q\rangle &\to& |i_L \rangle \otimes |A_a\rangle \nonumber \\
E_b |i_Q\rangle &\to& |i_L \rangle \otimes |A_b\rangle \nonumber \\
E_a |j_Q\rangle &\to& |j_L \rangle \otimes |A_a\rangle \nonumber  \\
E_b |j_Q\rangle &\to& |j_L \rangle \otimes |A_b\rangle
\end{eqnarray}
In other words, the final state
of $A$ must be the same for both codewords $|i_Q\rangle$ and
$|j_Q\rangle$, and depend only on the error syndrome $a$ or $b$.
This condition is clearly required in order to
recover an initial {\em arbitrary} superposition $\sum_i a_i |i_L\rangle$
({\it i.e.}, the ancilla must be in a tensor product
with the $k$ logical qubits after decoding). 
Conditions (\ref{eq_KL1}) and (\ref{eq_KL2}) then result
straightforwardly from
the orthogonality of the logical states $|i_L\rangle$ and $|j_L\rangle$,
and the conservation of scalar products by unitarity.
\par

The above considerations also apply to the quantum {\it erasure} channel
in which the position of the $e$ erroneous 
bits is {\em known}~\cite{bib_grassl}.
Note that conditions (\ref{eq_KL1}) and (\ref{eq_KL2}) 
obviously correspond to the case where the
errors are applied at $t$ {\it unknown}
positions in the codeword. Clearly, if the error-correcting code aims
at correcting for {\em erasures} only,
the error operators $E_a$ and $E_b$ differ from each other
by one-bit error operators at the {\em same} positions only.
Therefore, as the
product of two such $e$-erasure operators is another $e$-erasure
operator (a linear combination of the $E_a$'s), the necessary and sufficient
condition for erasure-correction becomes~\cite{bib_grassl}
\begin{eqnarray}
\label{eq_KL1bis}
\langle i_Q | E_a | i_Q \rangle &=&  
\langle j_Q | E_a | j_Q \rangle  \\
\label{eq_KL2bis}
\langle i_Q | E_a | j_Q \rangle &=& 0
\qquad\qquad {\rm for~}i\ne j
\end{eqnarray}
It results that an error-correcting code correcting $t$ errors (at unknown
positions) is equivalent to an $e$-erasure correcting code with $e=2t$.
This equivalence will be very useful in the following because
the quantum erasure channel is easier to treat using an entropic approach.
Before coming to the information-theoretic analysis of quantum 
error-correcting codes (Section~IV),
let us first analyze classical error correction in terms of entropies.
This will make the classical-quantum correspondence more transparent.

\section{Entropic condition for classical error/erasure correction}

Just like in the quantum case,
one can define two classes of classical noisy channels, depending
on the fact that the errors occur at known or unknown locations. In the
former case, the located errors are called {\em erasures},
and an erasure-correcting code is such that, if $e$ bits out of the $n$ bits
are ``erased'', it is possible to recover the encoded logical word from
the $n-e$ remaining bits only~\cite{bib_cover}. In the latter case of 
classical codes capable of correcting $t$ errors at unknown positions
in codewords of size $n$, all the $n$ bits of the corrupted
codewords must be used in the 
decoding operation. Exactly as for quantum codes,
it is easy to show that a classical code can correct $t$ errors at unknown
locations if and only if the same code can correct $e=2t$
erasures at known locations. The proof is as follows.
Let us consider two codewords of length $n$, $w_i$ and $w_j$,
and two error strings, $e_a$ and $e_b$ (the bits in a codeword
are flipped where the corresponding bits in the error string
are equal to 1). To be able to recover $t$ errors, we must have
\begin{equation}  \label{eq_KL3}
w_i \oplus e_a \ne w_j \oplus e_b
\end{equation}
for any two codewords and for
all possible error strings having $t$ bits (or fewer)
equal to one. Here,
$\oplus$ is the bitwise addition modulo 2 and $\ne$ means that the two strings
must differ by at least one bit.
A classical code correcting t errors must therefore be such that the
distance between any two codewords is larger than or equal to 
$2t+1$, since the error strings $e_a$ and $e_b$ can have at most $t$ bits
equal to one, implying that $e_a \oplus e_b$ can have at most $2t$ bits
equal to one. Now, in the case of codes capable of correcting
$e$ erasures, the positions of the bits equal to one in $e_a$ and
$e_b$ are identical, so that $e_c \equiv e_a \oplus e_b$ can have at most
$e$ (rather than $2e$) bits equal to one and is therefore another $e$-error
string just as $e_a$ or $e_b$. Thus, the condition for recovering
$e$ erasures is
\begin{equation}   \label{eq_KL4}
w_i \oplus e_c \ne w_j
\end{equation}
In other words, the distance between any two codewords
must only be larger than or equal to $e+1$. 
Obviously, Eq.~(\ref{eq_KL3}) parallels Eqs.~(\ref{eq_KL1}-\ref{eq_KL2}),
while Eq.~(\ref{eq_KL4}) parallels Eqs.~(\ref{eq_KL1bis}-\ref{eq_KL2bis}).
The resulting equivalence $e=2t$ will be important for our concern
because the entropic analysis is more adapted to erasure correction.
\par

\begin{figure}
\caption{Entropy diagram for a classical
erasure-correcting code. The input $X$ stands for the logical bits,
while the output $Y$ (the codewords) is split into
the erased bits $Y_e$ and the unchanged bits $Y_u$.
The condition for erasure-correction is
$H(X{\rm:}Y_e|Y_u)=0$, that is the entire information must
be found in the unchanged bits, $H(X{\rm:}Y_u)=I$. }
\vskip 0.25cm
\centerline{\psfig{figure=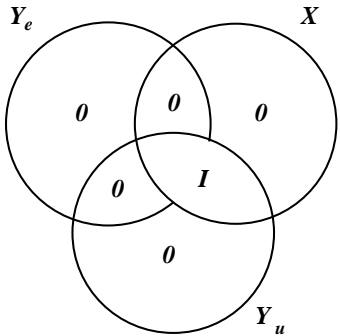,width=1.75in,angle=-90}}
\label{fig_class}
\vskip -0.25cm
\end{figure}

Let us shortly describe coding in the case of
a classical erasure channel~\cite{bib_cover}.
We consider encoding as a classical channel whose input $X$
is made of $k$ logical bits and output $Y$ is made of $n$ physical
bits (the codewords). We assume that the set of
logical words $x_i$ occur with probability $p_i$, so that the
entropy of the input $X$ is
\begin{equation}  \label{eq_entropyX}
H(X)=- \sum_i p_i \log p_i
\end{equation}
The input $X$ can be recorded (a classical variable can be ``cloned'') and
thus compared with the output $Y$.
As the encoding is reversible (it is a one-to-one mapping), the mutual
entropy is conserved through encoding, that is
\begin{equation} \label{eq_basiccondition}
I \equiv H(X{\rm:}Y)=H(X{\rm:}X)=H(X)
\end{equation}
where $I$ is defined as the mutual entropy (or information)
between input and output
that must be preserved in the classical erasure channel.
Let us assume that $Y$ is split
into $e$ erased bits, $Y_e$, and $n-e$ unchanged bits, $Y_u$.
(The position of the erased and unchanged bits is known.)
The condition for classical erasure correction is clearly
that the uncertainty of the input when the $n-e$ unchanged bits
are known vanishes, that is
\begin{equation}
H(X|Y_u)=0  \;.
\end{equation}
In other words, this means
that the $e$ bits can be erased without preventing the ability
of inferring the input $X$ from $Y_u$ without error. 
Since we have $H(X|Y)=H(X|Y_e Y_u)=0$ 
as a result of Eq.~(\ref{eq_basiccondition}), {\it i.e.}, it is obviously 
possible to infer $X$ from $Y\equiv Y_e Y_u$,
we obtain the basic entropic condition for classical error correction
\begin{equation}  \label{eq_classcond}
H(X{\rm:}Y_e|Y_u)=H(X|Y_u)-H(X|Y_e Y_u)=0
\end{equation}
Physically this expresses that, conditionally on the $n-e$ unchanged bits, no
information about $X$ is lost in the $e$ erased bits.
Classical coding works because the $n-e$ unaffected bits contain
the entire information $I$ about $X$, that is 
\begin{equation}  \label{eq_classcondbis}
H(X{\rm:}Y_u) = H(X{\rm:}Y) = I  \;,
\end{equation}
so that the $e$ bits that are erased are ``redundant''. 
Using the chain rule for Shannon mutual entropies,
\begin{equation}
H(X{\rm:}Y_e Y_u) = H(X{\rm:}Y_u)+H(X{\rm:}Y_e|Y_u)
\end{equation}
it is clear that Eq.~(\ref{eq_classcondbis}) is satisfied if and only if
the condition Eq.~(\ref{eq_classcond}) is satisfied.
In an erasure-correcting code, the $k$ bits of information
are thus distributed among the $n$ bits of $Y$ in such a way that
condition Eq.~(\ref{eq_classcond}) is satisfied
for any splitting of the $n$ bits into $e$ erased and $n-e$ unchanged bits.
The general classical entropy diagram corresponding to this situation
is represented in Fig~\ref{fig_class}. The condition for erasure
correction, Eq.~(\ref{eq_classcond}), appears in this Figure as
the vanishing entropy shared by $X$ and $Y_e$, but {\em not} by $Y_u$.
\par

Let us briefly show that, for a classical code, it is impossible
that {\it all} the patterns of $e$ ``erasable'' bits are independent of $X$, 
{\it i.e.}, do not contain some {\it redundant} information about
$X$. (This feature turns out to be possible for a quantum code,
as shown in Section~IV.) Suppose that we
could isolate two subparts of $Y$ {\it independent} of $X$, that is two
patterns of bits, say $Y_1$ and $Y_2$, such that
\begin{equation}   \label{eq_incompatible}
H(X{\rm:}Y_1)=H(X{\rm:}Y_2)=0
\end{equation}
Suppose also that, taken together, $Y_1$ and $Y_2$ provide the
entire information about $X$, that is
\begin{equation}
H(X{\rm:}Y_1 Y_2)=I
\end{equation}
This should be the case if we want to make a set of bits that
fully determines $X$ (such as $Y_u$) out of pieces that are independent
of $X$. We have
\begin{eqnarray}
H(X{\rm:}Y_1)+H(X{\rm:}Y_2) &=& H(X{\rm:}Y_1 Y_2) + H(X{\rm:}Y_1{\rm:}Y_2)
\nonumber \\
&=& I + H(Y_1{\rm:}Y_2) - H(Y_1{\rm:}Y_2|X)
\end{eqnarray}
Since the logical word $X$ fully determines any bit of the codeword $Y$, we
have $H(Y_1{\rm:}Y_2|X)=0$. Thus, the subadditivity of entropies,
$H(Y_1{\rm:}Y_2) \ge 0$, implies that
\begin{equation}
H(X{\rm:}Y_1)+H(X{\rm:}Y_2) \ge I
\end{equation}
which is incompatible with Eq.~(\ref{eq_incompatible}) if $I>0$.
One of the subpart ($Y_1$ or $Y_2$) must necessarily be correlated with $X$ 
(have a non-vanishing mutual entropy with $X$)
if the other one is independent of $X$.
Some pattern of $e$ ``erasable'' bits, including $Y_1$
or $Y_2$, will therefore be redundant (contain some information about
$X$ that is already in $Y_u$) as a consequence of strong subadditivity. 
\par

\section{Entropic condition for quantum error/erasure correction}

\subsection{Classical correspondence}

The above information-theoretic analysis can be straightforwardly
applied to the case of a quantum erasure-correcting code.
Here, the reference $R$ plays the role of the input $X$, while
$Q$ (the quantum codewords) replaces the output $Y$. We also
substitute the classical notion of Shannon mutual entropy (information)
between $X$ and $Y$ with the quantum notion of von Neumann mutual entropy
between $R$ and $Q$, and use the extension to the quantum regime
of the fundamental relations between Shannon entropies in a multipartite 
system~\cite{bib_neginfo,bib_measure,bib_channel} 
(see Appendix~\ref{sect_app}).
First, the mutual entropy between the logical words $L$ and $R$
is conserved through encoding (since it is unitary),
so that we have
\begin{equation}
I_q \equiv S(R{\rm:}Q)=S(R{\rm:}L)
\end{equation}
for the mutual entropy between the codewords $Q$ and $R$.
Here, $I_q$ can be seen as the ``quantum information'' 
(the entanglement with $R$) which must be preserved
in the quantum erasure channel.
As before, we assume that $Q$ is split into $Q_e$ (the $e$ erased qubits)
and $Q_u$ (the $n-e$ unchanged qubits). Just like in the classical case,
it is intuitively clear that entanglement is preserved at the condition that
the total mutual entropy with $R$ 
is found in the unaffected qubits, $Q_u$, that is
\begin{equation}
S(R{\rm:}Q_u) = S(R{\rm:}Q) = I_q
\end{equation}
in analogy with Eq.~(\ref{eq_classcondbis}).
Using the chain rule for the quantum mutual entropy
between $R$ and $Q\equiv Q_e Q_u$,
\begin{equation}
S(R{\rm:}Q_e Q_u) = S(R{\rm:}Q_u)+S(R{\rm:}Q_e|Q_u)
\end{equation}
we conclude that the condition for quantum erasure correction is
\begin{equation}   \label{eq_analclasscond}
S(R{\rm:}Q_e|Q_u)=0
\end{equation}
the straightforward analogue of Eq.~(\ref{eq_classcond}).
At this point, the parallel with classical erasure correction breaks
down because of a peculiar property of quantum entropies.
It is shown in Ref.~\cite{bib_measure} that the {\it ternary} mutual entropy
of any entangled tripartite system in a pure state vanishes
(see also Appendix~\ref{sect_app}).
In the case of interest here, the tripartite
system $RQ_e Q_u$ is in the pure state $|\psi_{RQ}\rangle$, so that
we have $S(R{\rm:}Q_e{\rm:}Q_u)=0$. As a consequence, we obtain
from Eq.~(\ref{eq_analclasscond})
the basic entropic condition for quantum erasure correction
\begin{equation}  \label{eq_quantcond}
S(R{\rm:}Q_e)=S(R{\rm:}Q_e|Q_u)+S(R{\rm:}Q_e{\rm:}Q_u)=0
\end{equation}
Physically, this expresses that the ``erased'' part of the codewords $Q_e$
must be {\em independent} of the reference $R$. This is very different from
the classical situation, where, in order to enable erasure correction,
the erased bits must by construction be correlated with $X$. In other
words, ``classical redundancy'' requires correlation of the $e$ redundant 
bits with $X$, while
``quantum redundancy'' is achieved {\em without} correlating
(or entangling) the erased qubits with $R$.
We will show later on that the above entropic condition,
Eq.~(\ref{eq_quantcond}), can be derived
more rigorously, using the property of strong subadditivity of quantum
entropies and the entropic condition for perfect
quantum error correction~\cite{bib_schum,bib_channel}.
\par

\subsection{Quantum loss of a noisy channel}

\begin{figure}
\caption{Entropy diagram summarizing the entropic relations between the
entangled systems $Q'$ (quantum channel), $R'$ (reference), and $E'$
(environment) after decoherence (see also Ref.~\protect\cite{bib_channel}). }
\vskip 0.25cm
\centerline{\psfig{figure=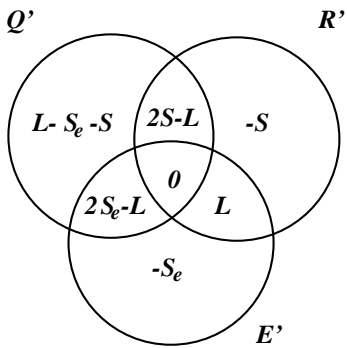,width=1.75in,angle=-90}}
\label{fig_diagram}
\vskip -0.25cm
\end{figure}

As explained in Section~II, we assume that the codewords sent on the
quantum noisy channel suffer an arbitrary decoherence due to the
environment $E$, that is $U_{QE}$ is an arbitrary unitary transformation.
(We do not restrict ourselves to a quantum erasure channel for the moment.)
After such an an arbitrary environment-induced decoherence,
the joint system $R'Q'E'$ is in the state $|\psi_{R'Q'E'}\rangle$ given
by Eq.~(\ref{eq_Uenvir}). The corresponding quantum entropy diagram 
is represented in Fig.~\ref{fig_diagram} (as mentioned earlier,
the primes refer to the systems {\em after} decoherence). 
As shown in Ref.~\cite{bib_channel},
it depends on three parameters, $S=S(R')=S(R)$, the entropy
of the reference $R$ (which is also equal to the entropy of $Q$
before decoherence),
$S_e=S(E')$, the entropy of the environment after decoherence, 
and\footnote{Since the total
system $R'Q'E'$ is in a pure state after decoherence,
{\it i.e.}, $S(R'Q'E')=0$, its Schmidt decomposition
implies $S(R'Q')=S(E')$ and $S(E'Q')=S(R')$, 
resulting in the last relation in Eq.~(\protect\ref{eq_loss}). }
\begin{eqnarray}  \label{eq_loss}
L&=&S(R'{\rm:}E'|Q') \nonumber \\
&=&S(R'Q')+S(E'Q')-S(Q')-S(R'Q'E') \nonumber \\
&=&S(E')+S(Q)-S(Q') \;,
\end{eqnarray}
the {\em loss} of the channel (following the terminology of Shannon
theory~\cite{bib_ash}).
The quantum loss $L$ can be shown to be the analogue of the loss
in a classical noisy channel, and thus can be written as a quantum 
conditional mutual entropy, {\it i.e.}, the quantum mutual entropy
between $R'$ and $E'$, conditionally on $Q'$~\cite{bib_channel}.
\par

The loss $L$ has a simple physical interpretation in the case of a 
{\em classical} noisy channel: it corresponds
to the entropy of the input $X$ of the
channel conditional on its output $Y$, {\it i.e.}, $L=H(X|Y)$,
thereby characterizing the unavoidable uncertainty in the decoding
operation (when inferring the input from the corrupted output).
Equivalently, it corresponds
to the mutual entropy between the input and the environment,
conditional on the output, {\it i.e.}, $L=H(X{\rm:}E|Y)$.
That is, for a given output, $L$ measures
the information about the input that has been irrecoverably lost in
correlations with the environment. If $X$ corresponds to encoded codewords
and $Y$ to corrupted ones due to a particular error source,
the condition $L=0$ must be satisfied for the error-correcting code
to preserve the codewords against classical noise~\cite{bib_ash}.
\par

\begin{figure}
\caption{Entanglement between $Q'$, $R'$, and $E'$ in a lossless ($L=0$)
quantum  channel. The quantum system $Q'$ is entangled ``separately''
with $R'$ and $E'$ (see also Ref.~\protect\cite{bib_channel}).}
\vskip 0.25cm
\centerline{\psfig{figure=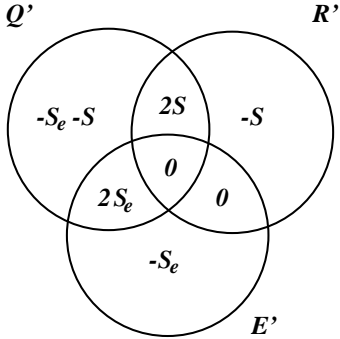,width=1.75in,angle=-90}}
\label{fig_L=0}
\vskip -0.25cm
\end{figure}

In Ref.~\cite{bib_channel}, it is shown that the same
interpretation holds for the quantum loss $L$, substituting
the classical notion of mutual information between $X$ and $E$
(conditional on $Y$)
with the quantum notion of von Neumann mutual entropy between $R'$ and $E'$
(conditional on $Q'$). The reference $R$ ($=R'$)
plays the role of the input $X$, while $Q'$ replaces the output $Y$.
Accordingly, it is expected that a vanishing quantum loss corresponds to a
situation where decoherence can be entirely eliminated using
a quantum code. Indeed,
\begin{equation}  \label{eq_L=0}
L=S_e+S-S(Q')=0
\end{equation}
is a necessary and sufficient condition
for the existence of a perfect quantum error-correcting code,
as proven recently by Schumacher and Nielsen~\cite{bib_schum}.
In Fig.~\ref{fig_L=0}, the entropy diagram of $R'Q'E'$ 
is represented in the case where this condition is achieved.
It appears that, when $L=0$, the state of
$Q'$ becomes entangled {\it separately} with the environment
(``bad'' entanglement)
and the reference (``good'' entanglement), allowing this ``bad'' entanglement
to be transfered to an ancilla (the $n-k$ check qubits)
while recovering only the ``good'' one.
This transfer of entanglement, requiring a local action on $Q$ only
(not on $E$), can be seen as a measurement of the error syndrome
(the ancilla becoming entangled with $E$) leaving the original state intact.
The fact that Eq.~(\ref{eq_L=0}) is a necessary condition can be understood
simply by noticing that the loss can never decrease
by processing $Q'$ through a subsequent channel, for example in the decoding
operation~\cite{bib_channel}.
Denoting the loss after decoherence by $L_1$ and the overall
loss (after decoherence and decoding) by $L_{12}$, one has
\begin{equation}
0\le L_1 \le L_{12}
\end{equation}
showing that $L_1=0$ is necessary for having $L_{12}=0$,
that is for perfectly recovering decoherence by decoding.
\par

Unlike in the classical case, it is possible to rewrite the quantum loss
as a function of $E'$ and $R'$ only, exploiting a purely quantum
feature of entropies in a tripartite system. 
As mentioned earlier, an important
consequence of $S(R'Q'E')=0$ is that the quantum {\it ternary} mutual entropy
vanishes, that is
\begin{eqnarray}
S(R'{\rm:}E'{\rm:}Q') &=& S(R'{\rm:}E')- S(R'{\rm:}E'|Q') \nonumber \\
&=& S(R')+S(Q')+S(E')-S(R'Q') \nonumber \\
&~& -S(R'E')-S(Q'E')+S(R'Q'E')   \nonumber \\
&=& 0
\end{eqnarray}
As a result, the quantum loss can be expressed as
\begin{equation}
L=S(R'{\rm:}E')
\end{equation}
Therefore, a necessary and sufficient condition for perfect
error correction is that the reference and the environment
are statistically {\it independent} ($L=0$). This condition
relates entropies {\em after} decoherence, and thus allows us to
check that, for a given code and after a specific interaction
with the environment, decoherence can be recovered by decoding. 
As far as quantum coding is concerned, it is more useful
to derive an entropic relation involving only the
reference $R$ and the codewords $Q$ {\em before}
unitary interaction with the environment,
using some error model [{\it cf.} Eq.~(\ref{eq_quantcond})].
\par

\subsection{Upper bound on the quantum loss}

As before, we consider now an explicit error model in which the decoherence
involves $e$ qubits at know locations, {\it i.e.}, the case of $e$ erasures.
The component $Q_e$ (of $e$ qubits) of the codeword 
interacts with $E$ while the rest $Q_u$ (of $n-e$ qubits)
remains unchanged by the interaction. Accordingly, the
unitary transformation describing such an error model
is $U_{RQE}= 1_{R} \otimes 1_{Q_u} \otimes U_{Q_e E}$.
This results in the conservation rule for the mutual 
entropy (see Appendix~\ref{sect_app}),
\begin{equation} \label{eq_expr1}
S(R' {\rm:} Q_e' E' ) = S(R {\rm:} Q_e E) = S(R{\rm:}Q_e)
\end{equation}  
where we made use of the fact that $E$ is initially in a pure state,
{\it i.e.}, $S(E)=0$. 
This entropy can also be expressed as
\begin{equation}  \label{eq_expr2}
S(R' {\rm:} Q_e' E') = S(R'{\rm:}E') + S(R'{\rm:}Q_e' | E')
\end{equation}
by use of the chain rule for quantum mutual entropies.
Using the strong subadditivity of quantum entropies,
\begin{eqnarray}
\lefteqn{S(R'{\rm:}Q_e' | E') = S(R' E')+S(Q_e' E')} 
\hspace{7truecm}\nonumber \\
-S(E')-S(R'Q_e'E') \ge 0
\end{eqnarray}
and denoting by
\begin{equation}  \label{eq_M}
M=S(R{\rm:}Q_e)
\end{equation}
the initial mutual entropy (or mutual entanglement)
between the reference $R$ and the {\it erased} subpart $Q_e$
of the codeword, Eqs. (\ref{eq_expr1}) and (\ref{eq_expr2})
yield an upper bound on the loss $L$:
\begin{equation}   \label{eq_LleM}
0 \le L \le M
\end{equation}
Consequently, if the mutual entanglement $M$ (initial mutual entropy
between $R$ and $Q_e$) is zero,
\begin{equation}   \label{eq_M=0}
M=S(R{\rm:}Q_e)=0 \;,
\end{equation}
then the loss $L=S(R'{\rm:}E')$ vanishes, allowing for perfect erasure
correction. In other words,
the statistical independence ($M=0$) between the reference $R$ and
the erased part of the codeword $Q_e$ is a {\it sufficient}
condition for perfect erasure correction, as anticipated
in Eq.~(\ref{eq_quantcond}). Note that this condition
must hold for any pattern of $e$ erased qubits among the $n$ qubits,
a constraint which implies the quantum Singleton bound (see Section~V).
\par

\begin{figure}
\caption{Entropy diagram for a quantum erasure-correcting code. It
characterizes the combined system $RQ\equiv RQ_eQ_u$
before decoherence when the condition
for perfect error correction $S(R{\rm:}Q_e)=0$ is fulfilled. The two
parameters are $S(R)=k$ and $S(Q_e)=s$.}
\vskip 0.25cm
\centerline{\psfig{figure=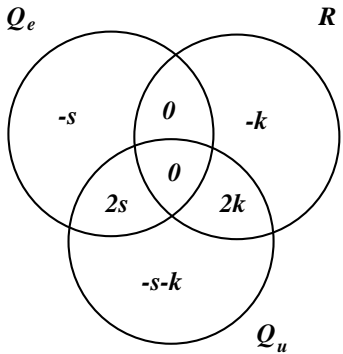,width=1.75in,angle=-90}}
\label{fig_quantumcode}
\vskip -0.25cm
\end{figure}

The physical content of the entropic condition,
Eq.~(\ref{eq_M=0}), is the following.
The reduced density matrix 
$\rho_{RQ_e}= {\rm Tr}_{Q_u}|\psi_{RQ}\rangle \langle \psi_{RQ}|$
obtained by tracing the state of $RQ$,
{\it i.e.}, Eq.~(\ref{eq_wavefunc}),
over $Q_u$ (ignoring the $n-e$ unchanged qubits) before decoherence 
must represent two {\it independent}
systems: the $k$ qubits of the reference $R$ and the $e$ erased qubits
$Q_e$ of the quantum system. The latter $e$ qubits can then be ``erased''
without interfering with $R$ in the sense that the $n-e$ remaining qubits
retain all the entanglement with $R$. The general entropy diagram of
the joint state of the system $RQ \equiv R Q_e Q_u$ before decoherence
is shown in Fig.~\ref{fig_quantumcode} (to be compared with
Fig.~\ref{fig_class} for a classical code).
For a code to  protect an arbitrary mutual entropy between $Q$ and $R$
(or an arbitrary state for $Q$), the
above condition $M=0$ must clearly be satisfied for the worst
case in which the amplitudes $a_i$ in Eq.~(\ref{eq_wavefunc}) are all equal
($|a_i|^2=2^{-k}$), that is in the case where $Q$ and $R$ ``saturate''
their entropy:
\begin{equation}   \label{eq_saturate}
S(Q)=S(R)=k
\end{equation}
(Note that, since $d_Q > d_L$, the entropy is limited by the
size of the Hilbert space of $R$.)
We will thus only consider this case in the following (it is important
when deriving the Singleton bound on quantum codes). Because of the
two constraints Eqs.~(\ref{eq_M=0}) and (\ref{eq_saturate}),
the ternary entropy diagram for $R Q_e Q_u$ depends on a single
unknown parameter, $s=S(Q_e)$, the entropy of the erased 
qubits\footnote{The entropy diagram for a general tripartite system
in a pure state depends on three parameters.}.
In view of Fig.~\ref{fig_quantumcode}, we see that
the $e$ qubits of $Q_e$ are ``superfluous'', as they
do not yield any information about $R$ (no mutual entanglement with $R$)
and are thus
unnecessary for recovering the original logical state. In this sense,
they constitute ``redundant'' quantum information since the total
mutual entropy $2k$ is found between $R$ and $Q_u$.
The unchanged qubits $Q_u$ are entangled {\em separately} with the
reference $R$ (``useful'' entanglement $2k$ that must be preserved
by the code) and with the erased qubits $Q_e$ (``useless'' entanglement),
so that any action on $Q_e$ due to an environment $E$ can only
transfer this ``useless'' entanglement to $E$
but leaves the ``useful'' entanglement unchanged. Indeed,
if the entropy diagram Fig.~\ref{fig_quantumcode} is achieved, 
then Eq.~(\ref{eq_LleM}) implies that
any interaction between $Q_e$ and $E$ necessarily
results in an entropy diagram such as the one
depicted in Fig.~\ref{fig_L=0}, where $Q'$ is entangled {\it separately}
with $E'$ and $R'$ ({\it i.e.}, $L=0$),
guaranteeing that one can undo decoherence
by applying an appropriate decoding. 

\subsection{Example}

\begin{figure}
\caption{Quantum entropy diagram of the combined system $RQ_eQ_u$
before decoherence for the 5 bit quantum code ($n=5$, $k=1$, $e=2$)
\protect\cite{bib_laflamme,bib_bdsw}.}
\vskip 0.25cm
\centerline{\psfig{figure=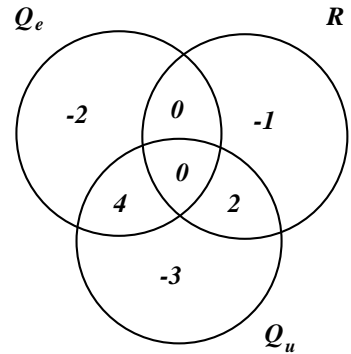,width=1.75in,angle=-90}}
\label{fig_5bitcode}
\vskip -0.25cm
\end{figure}

As an illustration, we show in Fig.~\ref{fig_5bitcode}
the entropy diagram in the case of a 5 qubit code ($n=5$) encoding
$k=1$ logical qubit with $t=1$, {\it i.e.}, allowing up to $e=2$ 
erasures~\cite{bib_laflamme,bib_bdsw}. The full mutual
entanglement of 2 bits is found between $R$ and $Q_u$, while the 2 erased
qubits $Q_e$ are independent of $R$. We have $S(R)=1$, $S(Q_u)=3$,
and $S(Q_e)=2$, so that each subsystem has the maximum allowed entropy
for its Hilbert space ($R$, $Q_u$, and $Q_e$ are made of 1, 3, and 2
qubits, respectively). Note that the 4 qubit code ($n=4$) encoding
$k=2$ logical qubits and correcting $e=1$ 
erasure~\cite{bib_vaidman,bib_grassl} corresponds
in fact to the same entropy diagram with $R$ playing the role of
$Q_e$ and conversely. Indeed, $R$ has then an entropy of 2 bits
and shares a mutual entropy of 4 bits with $Q_u$ (which then contains the
full information about the 2 encoded qubits). This mutual entanglement
is preserved against erasure of 1 qubit since $Q_e$ is independent of $R$.
\par

\begin{figure}
\caption{Classical entropy diagram of $X Y_e Y_u$ for the
$n=5$, $k=2$, $e=2$ classical linear code defined in the text.}
\vskip 0.25cm
\centerline{\psfig{figure=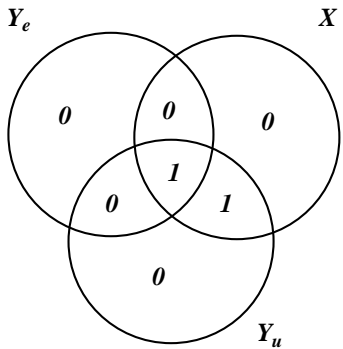,width=1.75in,angle=-90}}
\label{fig_classcode}
\vskip -0.25cm
\end{figure}

Let us finally compare this ``quantum redundancy'' in the 5 qubits code
with the entropic diagram characterizing a classical code.
As explained in Section~III,
in a classical code the information ($k$ bits) is distributed among
the $n$ bits which are then correlated with the input $X$
in a specific way [so that Eq.~(\ref{eq_classcond}) is satisfied].
Ignoring the $n-e$ unchanged bits ($Y_u$)
leaves $e$ bits ($Y_e$) that are {\em redundant} [the total information
is in $Y_u$, as implied by Eq.~(\ref{eq_classcondbis})],
but {\em correlated} with the input $X$,
in contrast with the quantum case. 
The entropy diagram corresponding to a simple classical code
is illustrated in Fig.~\ref{fig_classcode} (to be compared
with Fig.~\ref{fig_5bitcode}).
We consider a simple linear code with $n=5$, $k=2$, $e=2$,
defined in Ref.~\cite{bib_cleve}
\begin{eqnarray}
00 &\longrightarrow& 00000 \nonumber \\
01 &\longrightarrow& 01110 \nonumber \\
10 &\longrightarrow& 10101 \nonumber \\
11 &\longrightarrow& 11011 
\end{eqnarray}
such that the distance between any two codewords is 3 or larger.
We assume that the first and the last bit are erased ($Y_e$),
the three other ones
being unchanged ($Y_u$), and show the entropy diagram in the case where
the logical words $00$ to $11$ are equiprobable [the entropy
in Eq.~(\ref{eq_entropyX}) is maximum].
The full information is found in $Y_u$, $H(X{\rm:}Y_u)=2$~bits, but the erased
bits are partially correlated with $X$, $H(X{\rm:}Y_e)=1$~bit.
Classical redundancy necessarily implies that the erased bits 
contain part of the information that is duplicated.
The situation is thus quite different in a quantum code:
Eq.~(\ref{eq_classcond}) is replaced by its quantum counterpart
Eq.~(\ref{eq_analclasscond}), with the same physical interpretation,
but the latter equation then
implies the simpler condition~(\ref{eq_M=0}) as a consequence of
the property that the quantum ternary mutual entropy vanishes 
for a pure state. Such a possibility
to achieve ``weak'' cloning through coding (in the sense
that the full information is in $Q_u$ without correlating $Q_e$
with the reference $R$) is purely quantum and
suggests an interesting interpretation of quantum coding,
as explained in Section~VI.

\section{Singleton bound on quantum codes}

The above entropic considerations provide a simple way
to derive the quantum analogue of the Singleton bound on error-correcting
codes, obtained recently by Knill and Laflamme~\cite{bib_knilaf}.
For a classical code, the Singleton bound (see, {\it e.g.},
\cite{bib_lint}) states that
the number of logical bits $k$ than can be encoded in a code
of length $n$ recovering $e$ erasures is such that
\begin{equation}   \label{eq_class_singleton}
k \le n - e
\end{equation}
Of course, for a classical code recovering $t$ errors (at unknown locations),
the Singleton bound becomes
\begin{equation}
k \le n - 2t
\end{equation}
as a consequence of the equivalence between codes correcting $t$ errors
and $e=2t$ erasures. In order to derive the quantum analogue of this bound, 
we consider the joint state of the system 
$RQ = R Q_e Q_u$ before decoherence.
For a quantum code to protect an arbitrary entanglement between $Q$ and $R$,
the entropic condition $M=S(R{\rm:}Q_e)=0$ must is satisfied for the
worst case of maximum entanglement, that is in the case where $Q$ and $R$
``saturate'' their entropy $S(Q)=S(R)=k$.  Assume for the moment
that the code is such that $S(Q_e)=e$, {\it i.e.} that the erased
qubits have the maximum entropy allowed by the dimension
of the Hilbert space of $Q_e$. Then,
the condition $M=0$ is clearly satisfied if tracing over the $n-e$
qubits associated with $Q_u$ yields a reduced density matrix for $R Q_e$
that saturates its quantum entropy
\begin{equation}
S(R Q_e) = S(R)+S(Q_e)-S(R{\rm:}Q_e) = k+e
\end{equation}
This corresponds to the case $s=e$ in Fig.~\ref{fig_quantumcode}.
Since $RQ_eQ_u$ is in a pure state [{\it i.e.}, $S(RQ_eQ_u)=0$], one has
$S(RQ_e)=S(Q_u)$ by Schmidt decomposition.
Expressing that the quantum entropy of $Q_u$ is bounded from
above by the logarithm of the dimension of its Hilbert space,
that is $S(Q_u) \le n-e$, one gets the inequality
\begin{equation}  \label{eq_KLerasure}
k \le n - 2 e 
\end{equation}
which is the Singleton bound for quantum
erasure-correcting codes. Making use of the equivalence between
codes correcting errors of $t$ qubits and the erasure of $e=2t$ qubits,
we get the Singleton bound for quantum error-correcting codes 
(proven in Ref.~\cite{bib_knilaf} for $k=t=1$)
\begin{equation}   \label{eq_KLerror}
k \le n - 4 t 
\end{equation}
This condition must be satisfied by
any quantum code (including degenerate codes). Mathematically,
Eq.~(\ref{eq_KLerasure}) expresses thus that it is necessary to trace over
at least half of the $n+k$ qubits constituting the total entangled
state $|\psi_{RQ}\rangle$ in order to open the possibility of having
$k+e$ independent remaining qubits (that is which saturate their entropy),
thereby allowing error correction. Eq.~(\ref{eq_KLerasure})
suggests an interpretation of quantum coding in terms of 
a ``weak'' cloning, as explained in the next Section.
\par

\begin{figure}
\caption{Schematic representation of the two different splittings of $Q$
into $Q_e Q_u$ or $Q_e' Q_u'$ which are used in the derivation of the
quantum Singleton bound.}
\vskip 0.25cm
\centerline{\psfig{figure=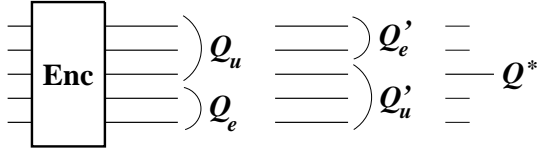,width=2.8in,angle=-90}}
\label{fig_singleton}
\vskip -0.25cm
\end{figure}

The above derivation was based on the assumption that the erased
qubits have a maximum entropy, {\it i.e.}, $S(Q_e)=e$. This is true
for example in the case of the 5 qubit code shown in 
Fig.~\ref{fig_5bitcode}. However, we need to prove
Eq.~(\ref{eq_KLerasure}) in full generality,
without recourse to this assumption.
In general, it is possible to have $S(R{\rm:}Q_e)=0$ with $S(Q_e) < e$;
this is the case for example when one (or more) of the physical
qubits is always 0 (non-optimal code). Suppose that the condition
for erasure correction
$S(R{\rm:}Q_e)=0$ is satisfied for some pattern of $e$ erased qubits,
so that the remaining part of the codeword $Q_u$ 
retains the ``full'' entanglement,
$S(R{\rm:}Q_u)=2k$. The central point in deriving the quantum
Singleton bound is that this entropic condition must be fulfilled for
{\it any} pattern of $e$ erased qubits among the $n$ qubits. Therefore, one
can choose for example another pattern of $e$ qubits within
the $n-e$ qubits that constitute $Q_u$, and check that
they have also a vanishing mutual entropy with $R$.\footnote{This
 implies the simple constraint that the dimension
 of $Q_u$ must be larger that the dimension of $Q_e$, that is
 $e<n-e$. This inequality, {\it i.e.}, the Singleton bound for $k=1$,
 also results straightforwardly from
 the no-cloning theorem (see Section VI).}
Let us denote the $e$ erased qubits in this second check
by $Q_e'$, so that $Q_u$ is divided into $Q_e'$ and
$Q^*$ (the $n-2e$ remaining qubits) as shown in Fig.~\ref{fig_singleton}.
The corresponding entropic condition is thus $S(R{\rm:}Q_e')=0$.
Conversely, the unchanged qubits $Q_u' \equiv Q_e Q^*$ in this second check
must also retain the full mutual entanglement with $R$, {\it i.e.},
$S(R{\rm:}Q_u')=2k$. This implies that,
while the $e$ qubits of $Q_e$ are independent of $R$, they must recover
the total mutual entanglement $2k$ with $R$ when supplemented only
with the $n-2e$ qubits of $Q^*$. These two opposite constraints
must be satisfied simultaneously,
which gives rise to the quantum Singleton bound.
\par

In order to prove this bound, we first
calculate a lower bound on the entropy of $Q^*$. 
Using the fact that $RQ = RQ_eQ_e'Q^*$ is in a pure state and
the independence between $Q_e$ and $R$, we have
\begin{eqnarray}
S(Q_u)&=&S(Q_e'Q^*) \nonumber \\
      &=&S(RQ_e) \nonumber \\
      &=&S(R)+S(Q_e)-S(R{\rm:}Q_e) \nonumber \\
      &=&k+S(Q_e)
\end{eqnarray}
Then, the property of subadditivity of quantum entropies
\begin{equation}
S(Q_e'Q^*) \le S(Q_e') + S(Q^*)
\end{equation}
implies the inequality
\begin{equation}  \label{eq_together1}
k + S(Q_e) - S(Q_e') \le S(Q^*)
\end{equation}
By the same token, given the independence between $Q_e'$ and $R$,
we can calculate the entropy of $Q_u'$,
\begin{eqnarray}
S(Q_u')&=&S(Q_eQ^*) \nonumber \\
       &=&S(RQ_e') \nonumber \\
      &=&S(R)+S(Q_e')-S(R{\rm:}Q_e') \nonumber \\
      &=&k+S(Q_e')
\end{eqnarray}
and make use of subadditivity
\begin{equation}
S(Q_eQ^*) \le S(Q_e) + S(Q^*)
\end{equation}
to obtain
\begin{equation}  \label{eq_together2}
k + S(Q_e') - S(Q_e) \le S(Q^*)
\end{equation}
Finally, combining Eqs.~(\ref{eq_together1}) and (\ref{eq_together2}) provides
a lower bound for $S(Q^*)$:
\begin{equation} \label{eq_bound_k}
k \le S(Q^*)
\end{equation}
This bound is equivalent to $S(Q_eQ_e'|R) \ge 0$, and has the following
interpretation. Even though $Q_e$ and $Q_e'$ are both independent
of $R$, the combined system $Q_eQ_e'$ will generally be entangled  
with $R$ (with a mutual entropy between 0 and $2k$). However,
in contrast with the entanglement between $Q_u$ (or $Q_u'$) and $R$,
the entropy of $Q_eQ_e'$ conditional on $R$ cannot become negative
because of the opposite constraints on $S(Q_e) - S(Q_e')$ from
Eqs.~(\ref{eq_together1}) and (\ref{eq_together2}).
The quantum Singleton bound is obtained simply by noticing
that $S(Q^*)$ is bounded from above by the dimension of the Hilbert
space of $Q^*$, that is
\begin{equation}
S(Q^*) \le n-2e
\end{equation}
The latter equation together with Eq.~(\ref{eq_bound_k})
completes the proof of Eqs.~(\ref{eq_KLerasure})
and (\ref{eq_KLerror}).

\section{Discussion and conclusion}

Before concluding, let us discuss the relation between the quantum no-cloning
theorem and quantum erasure-correcting codes.
The main point is to note that, if we can erase $e$ qubits
while being able to recover the codeword, it means that the $n-e$ remaining
qubits contain all the information, so that the $e$ qubits apparently
contain a (partial) duplication of the logical word. Clearly,
it is forbidden to erase half (or more) of the $n$ qubits, since then
the two halves of the codeword could be mapped on the logical word,
enabling quantum cloning. Thus, one must have $n - 2e >0$ (a constraint
equivalent to the condition
that the dimension of $Q_u$ must exceed the dimension of $Q_e$). 
However, Eq.~(\ref{eq_KLerasure}) is actually more restrictive, implying that
it is possible to quantify the impossibility of cloning (to be more precise
than a yes-or-no theorem). We will see that the Singleton bound expresses
that a ``weak'' cloning is allowed, up to a certain extent.
Let us define the number of clones (the {\em fractional} number of
copies of the logical word) as:
\begin{equation}
N_c \equiv { e \over n-e}
\end{equation}
where the $n-e$ qubits constitute the ``original'' (necessary to fully recover
the logical state) while the $e$ erased qubits make the 
partial\footnote{This concept of ``partial'' cloning is 
unrelated to the notion of ``approximate'' cloning introduced in
Ref.~\cite{bib_bh}. There, a universal quantum-cloning machine
is used that has two outputs being an {\it approximate} copy of the
input. This can be viewed as two channels sharing the same input
but necessarily characterized both by a {\it non-vanishing} quantum loss,
{\it i.e.}, the fidelity of both copies is not one.
In our case, we have two {\it lossless} channels: $L\to Q_u$ and $L\to
Q_u'$. However, the outputs unavoidably share a common piece
$Q^*$ which cannot be reduced to zero for 
$S(R{\rm:}Q_u)=2k$ and $S(R{\rm:}Q_u')=2k$ to hold simultaneously.}
clone.
It is easy to see from Eq.~(\ref{eq_KLerasure}) that
the fractional number of clones is restricted to the range
\begin{equation}  \label{eq_weakclone}
0 \le N_c \le { n-k \over n+k }
\end{equation}
This somehow extends the standard no-cloning theorem
to ``weak'' ($N_c <1$) cloning. The no-cloning theorem~\cite{bib_noncloning}
states that it is forbidden to make one full clone, {\it i.e.}, $N_c\ne 1$,
while Eq.~(\ref{eq_weakclone}) provides an upper bound on weak cloning.
In the limiting case where $n=k$, the number of clones is strictly
zero. This simply means that, if the codewords span the full $2^n$-dimensional
Hilbert space ({\it i.e.}, if no coding is actually used), 
then no cloning at all is achieved ($N_c=0$). The same is obviously
true for a classical code, since, using Eq.~(\ref{eq_class_singleton}),
the equivalent condition on ``weak'' cloning is $0 \le N_c \le (n-k)/k$.
When quantum coding uses only part of the Hilbert space,
{\it i.e.}, the space of
codewords is some $2^k$-dimensional subspace of the full space ($k<n$),
the logical states may then be viewed as partially cloned by the
encoding process,
the fractional number of clones being limited to $(n-k)/(n+k)$.
The latter increases as a smaller subspace is used ($k$ decreases), and
tends to one (full cloning) when $n/k \to \infty$.
The case $k=0$ corresponds to perfect cloning of a {\it fixed} 
({\it i.e.}, non-arbitrary) pure state. 
Therefore, whatever the apparent ``replication factor'' $n/k$ 
of the logical words achieved by the encoding process,
the allowed number of clones $N_c < 1$ (for a non-vanishing $k$).
For a classical code, however, no such limit exists on the number of clones,
as $N_c \to n/k$ when $n/k \to \infty$.
\par

We have shown that some new insight into quantum coding can be gained
by use of an information-theoretic approach paralleling the one
used to describe classical coding. Such an analysis displays explicitly
the similarities between classical and quantum codes,
but also emphasizes the major differences. The entropic condition
for a quantum erasure-correcting code is that the quantum mutual entropy
between a reference and the erased part of the codeword is
vanishing prior to decoherence. Such a statistical independence between
the reference and the erased qubits (interacting with the environment)
guarantees that the entanglement of the logical word with respect to
this reference is preserved by the quantum code. This is to be compared
with the corresponding entropic condition for a classical erasure-correcting
code, {\it i.e.}, that the mutual information between the logical bits and
the erased bits of the codewords, {\em conditional} on the remaining
unchanged bits of the codewords, is vanishing. Such a classical condition,
however, does {\em not} imply that the erased bits are independent
of the logical bits. On the contrary, there must be correlations between them,
and this duplication (or ``cloning'') of classical information
is at the heart of classical codes. Such a classical redundancy
has no quantum counterpart, as a consequence of the purely quantum
property that the {\em ternary} mutual entropy vanishes for any
entangled tripartite system in a pure state. In a quantum code,
only a ``weak'' cloning is achieved, up to the extent allowed by
the quantum Singleton bound, so that
the erased qubits are {\it unentangled} with the reference although the
entire codeword remains entangled with it. This reflects a major
difference between classical and quantum coding.

\acknowledgements
We acknowledge C. Adami and J. Preskill for very useful discussions.
We thank the organizers of the ITP program on Quantum Computers
and Decoherence for their invitation in Santa Barbara,
where this work has been performed.
This research was supported in part by the National Science Foundation
under Grant Nos. PHY 94-12818, PHY 94-20470 and PHY 94-07194,
and by a grant from DARPA/ARO through the QUIC Program
(\#DAAH04-96-1-3086).

\appendix
\section{Information-theoretical framework for quantum entropies}
\label{sect_app}

Classical information theory is centered on Shannon
entropies (see, {\it e.g.}, \cite{bib_ash,bib_cover}).
A random variable $X$, distributed according to
the probability distribution $p_i$, is characterized by the
Shannon entropy
\begin{equation}  \label{eq_app1}
H(X) \equiv -\sum_i p_i \log p_i
\end{equation}
The Shannon entropy $H(X)$ measures the {\it uncertainty} of $X$ (it
vanishes if the distribution is peaked, {\it i.e.}, if the value
of $X$ is perfectly known).
When considering two random variables $X$ and $Y$, described in general by
the joint probability distribution $p_{i,j}$, one can define
several entropies. First, one has
the joint entropy $H(XY)$, based on $p_{i,j}$ 
in analogy with Eq.~(\ref{eq_app1}), which reflects the uncertainty of
$X$ and $Y$. Second, one defines the entropy of $X$ {\it conditional} on
$Y$, that is the entropy of $X$ when $Y$ is known (averaged over $Y$),
\begin{eqnarray}  \label{eq_app2}
H(X|Y) &\equiv& -\sum_{i,j} p_{i,j} \log p_{i|j} \nonumber \\
       &=& H(XY)-H(Y)
\end{eqnarray}
based on $p_{i|j}=p_{i,j}/p_j$, the conditional probability of $i$
knowing $j$. The equivalent definition holds for the conditional
entropy of $Y$ knowing $X$, {\it i.e.}, $H(Y|X)=H(XY)-H(X)$.
Finally, one defines the {\it mutual} entropy between $X$ and $Y$ as
\begin{eqnarray}  \label{eq_app3}
H(X{\rm:}Y) &\equiv& -\sum_{i,j} p_{i,j} \log p_{i{\rm:}j} \nonumber \\
&=& H(X)+H(Y)-H(XY)
\end{eqnarray}
where $p_{i{\rm:}j}=p_i p_j / p_{i,j}$ is the mutual probability
of $i$ and $j$.
It plays the role of a mutual {\it information} between $X$ and $Y$, that is
the information about $X$ that is conveyed by $Y$,
or the decrease of the entropy of $X$ due to knowledge of $Y$ (or conversely):
\begin{equation}
H(X{\rm:}Y)=H(X)-H(X|Y)=H(Y)-H(Y|X)
\end{equation}
\par

Consider now the information-theoretical description of a 
bipartite quantum system $XY$, where $X$ and $Y$ correspond to two
quantum variables or degrees of freedom ({\it e.g.}, the $z$-component
of a spin-1/2 particle). It is shown in Ref.~\cite{bib_neginfo,bib_measure}
that a quantum information-theoretical
formalism that parallels Shannon construction can be
defined that is based on the von Neumann entropy,
where probability distributions are replaced by density matrices, and
averages are changed into quantum expectation values.
We have for the quantum (von Neumann) entropy of $X$,
\begin{equation}  \label{eq_app4}
S(X)\equiv  - {\rm Tr}_X ( \rho_X \log \rho_X )
\end{equation}
where $\rho_X$ is the density matrix characterizing the state of $X$.
The density matrix $\rho_X$ is obtained by a partial trace over the
$Y$ variable, {\it i.e.}, $\rho_X={\rm Tr}_Y (\rho_{XY})$,
where $\rho_{XY}$ is the joint density matrix describing $XY$.
A similar definition holds for $S(Y)$, based on $\rho_Y$.
One defines then the quantum (von Neumann) {\it conditional} entropy,
\begin{eqnarray}  \label{eq_app5}
S(X|Y)&\equiv& - {\rm Tr}_{XY} (\rho_{XY} \log \rho_{X|Y}) \nonumber \\
&=& S(XY)-S(Y)
\end{eqnarray}
based on the {\it conditional}
density matrix $\rho_{X|Y}$ (defined in~\cite{bib_neginfo,bib_measure}).
The latter plays the role of a ``quantum'' conditional probability,
and witnesses the appearance of non-classical correlations in the
case of quantum entangled variables $X$ and $Y$. Indeed, it can be
shown that an eigenvalue of $\rho_{X|Y}$ can exceed one, and,
consequently, 
that the quantum conditional entropy $S(X|Y)$ can be {\it negative},
a fact related to quantum non-separability
(see~\cite{bib_neginfo,bib_measure,bib_bell}). For example, if $X$ and
$Y$ represent two entangled quantum bits in
a Bell state, we have $S(X|Y)=S(Y|X)=-1~$bit.
Since negative conditional entropies are forbidden in Shannon theory,
a negative value of $S(X|Y)$ obviously implies quantum non-separability,
the converse being not true. On the other hand,
if $\rho_{XY}$ describes a mixture of orthogonal product states (that
is a classical situation), $\rho_{X|Y}$ is then a diagonal matrix with the
$p_{i|j}$'s on its diagonal, and Eq.~(\ref{eq_app5}) reduces to its classical
counterpart, Eq.~(\ref{eq_app2}). The above definition, Eq.~(\ref{eq_app5}),
must therefore be viewed as a quantum {\it extension}
of the Shannon conditional entropy
in a way that incorporates quantum entanglement, while
including the classical conditional entropy as a special case.
\par

According to this, it is natural to define a quantum 
(von Neumann) {\it mutual} entropy,
\begin{eqnarray}  \label{eq_app6}
S(X{\rm:}Y) &\equiv& 
- {\rm Tr}_{XY} (\rho_{XY} \log \rho_{X{\rm:}Y} )\nonumber \\
&=& S(X)+S(Y)-S(XY) \nonumber \\
&=& S(X)-S(X|Y)=S(Y)-S(Y|X)
\end{eqnarray}
based on the {\it mutual}
density matrix $\rho_{X{\rm:}Y}$ (defined in~\cite{bib_neginfo,bib_measure}).
The interpretation is the same as in Shannon information theory, and
$S(X{\rm:}Y)$ is a symmetric quantity. 
Subadditivity of quantum entropies, {\it i.e.}, $S(XY) \le
S(X)+S(Y)$, implies that
$S(X{\rm:}Y)\ge 0$, just as for Shannon mutual entropies. 
However, in the case of quantum entangled
variables, $S(X{\rm:}Y)$ can reach twice the maximum
allowed value in Shannon theory~\cite{bib_neginfo,bib_measure}:
\begin{equation}
S(X{\rm:}Y)\le 2 \min [S(X),S(Y)]
\end{equation}
For instance, we have $S(X{\rm:}Y)=2$~bits between the members
of an EPR pair. Note that
$S(X{\rm:}Y)$ is not a {\it measure} of entanglement in the sense that
it can be non-zero for classical (separable) mixtures. It
does not necessarily exceed the classical upper bound,
$\min [S(X),S(Y)]$, for quantum entangled systems. Thus, $S(X{\rm:}Y)$
is rather a quantum mechanical extension of the usual Shannon
mutual entropy, which measures quantum as well as classical correlations.
Nevertheless, it plays an important role in the information-theoretic
description of quantum channel and error correction, as emphasized
throughout this paper (see also~\cite{bib_channel}).
\par

Consider a quantum system $X$ entangled with a reference $R$, so that
the joint system $RQ$ is in the pure state 
$\sum_i \sqrt{p_i} |i_R\rangle |i_X\rangle$.
The system $X$ is sent through a quantum channel (which does not act
on $R$). It is easy to see that,
if the channel (including error-correction) is ``perfect'',
{\it i.e.}, if the output $Y$ ends up in a joint pure
state (together with $R$) 
such that the quantum mutual entropy with $R$ is conserved,
then an {\it arbitrary} quantum state is preserved 
in the channel\footnote{The classical analogue of this property
is intuitive. If the input $X$ of a classical channel
is copied into a memory $M$ (so that $M$ is thus perfectly correlated
with $X$), the correlation between the output $Y$ and the memory $M$
reflects the ``quality'' of the channel. In particular, if the mutual entropy
between $Y$ and $M$ is equal to that between $X$ and $M$,
then the classical channel is perfect (lossless).}.
Indeed, as $R$ is unchanged, any joint state of $RY$ that is
characterized by $S(R{\rm:}Y)=S(R{\rm:}X)$ and $S(RY)=S(RX)=0$
is necessarily of the form
$\sum_i \sqrt{p_i} |i_R\rangle (U |i_X\rangle)$ where
$U$ is a {\it fixed} unitary transformation (it does {\it not} depend
on the $p_i$'s). Thus, up to a given change of basis, the output $Y$
is in the same entangled state with $R$. Projecting $R$ onto any pure state
shows that the corresponding pure state of $X$ (the ``relative'' state)
has been preserved, {\it i.e.}, $Y$ ends up in the same state.
\par

Another important property of the quantum mutual entropy 
$S(X{\rm:}Y)$ is that
it is conserved when the bipartite system $XY$ undergoes
a unitary transformation of the form $U_X \otimes U_Y$. Indeed, if
\begin{equation}
\rho_{XY}'= (U_X \otimes U_Y) \rho_{XY} (U_X \otimes
U_Y)^{\dagger}
\end{equation}
then $\rho_X'= {\rm Tr}_{Y'} (\rho_{XY}') = U_X \rho_{X} U_X^{\dagger}$ and 
similarly for $\rho_Y'$,
so that
\begin{equation}
S(X'{\rm:}Y')=S(X{\rm:}Y)
\end{equation}
follows from Eq.~(\ref{eq_app6}) together with
the conservation of von Neumann entropy under a unitary
transformation. 
In particular, any entangled system $XY$ that undergoes a local
operation separately on $X$ and $Y$ retains its initial entanglement
between $X$ and $Y$, {\it i.e.}, $S(X{\rm:}Y)$ is conserved.
This property is useful in the context of quantum
channels and quantum error correction.
\par

The quantum information-theoretical formalism defined above
can be generalized to multipartite systems, in analogy to the
Shannon construction~\cite{bib_neginfo,bib_measure}. 
The definition of the conditional (and mutual)
density matrices provides grounds for the quantum extension
of the usual algebraic relations between Shannon entropies
(see, {\it e.g.}, \cite{bib_ash,bib_cover}). 
The resulting framework for quantum information theory goes 
beyond classical correlations, {\it i.e.}, accounts for situation
where $n$ quantum variables are entangled, by allowing
conditional entropies to be negative.
Of course, it also includes Shannon theory as a special case.
Consider, for example, a tripartite quantum system $XYZ$.
First, one can write the {\it chain rule} for quantum entropies,
\begin{equation}
S(XYZ)=S(X)+S(Y|X)+S(Z|XY)
\end{equation}
One can also define the von Neumann {\it conditional mutual} entropy,
\begin{eqnarray}   \label{eq_app7}
S(X{\rm:}Y|Z)&=&S(X|Z)-S(X|YZ) \nonumber \\
&=& S(X|Z)+S(Y|Z)-S(XY|Z) \nonumber \\
&=& S(XZ)+S(YZ)-S(Z)-S(XYZ)
\end{eqnarray}
which reflects the quantum mutual entropy between $X$ and $Y$, when $Z$ is
known. Eq.~(\ref{eq_app7}) follows the definition of a mutual entropy,
Eq.~(\ref{eq_app6}), but with all entropies being conditional on $Z$.
Just like with classical entropies, the property of {\it strong
subadditivity} of quantum entropies holds, that is $S(X{\rm:}Y|Z)\ge 0$.
Conditional mutual entropies are also used in the quantum analogue of
the chain rules for mutual entropies, that is
\begin{equation}
S(X{\rm:}YZ)=S(X{\rm:}Z)+S(X{\rm:}Y|Z)
\end{equation}
Finally, the relation between conditional and mutual entropies,
$S(X)=S(X|Z)+S(X{\rm:}Z)$ can be extended to a tripartite system,
that is
\begin{equation}
S(X{\rm:}Y)=S(X{\rm:}Y|Z)+S(X{\rm:}Y{\rm:}Z)
\end{equation}
so that we can split $S(X{\rm:}Y)$ into a conditional piece
and a mutual piece with $Z$.
The latter piece, $S(X{\rm:}Y{\rm:}Z)$, characterizes therefore the
{\it ternary} mutual entropy, {\it i.e.}, that piece of the mutual entropy
between $X$ and $Y$ that is also shared by $Z$.
All these relations between entropies can be understood very easily
using Venn diagrams~\cite{bib_neginfo,bib_measure,bib_channel}.
\par

Finally, an important property of {\it quantum} entropies
which has no classical counterpart, is that, 
for any entangled tripartite system $XYZ$ in a pure state,
the ternary mutual entropy vanishes~\cite{bib_measure}, {\it i.e.},
\begin{eqnarray}
S(X{\rm:}Y{\rm:}Z) &=& S(X)+S(Y)+S(Z)-S(XY) \nonumber \\
& &-S(XZ)-S(YZ)+S(XYZ) \nonumber \\
&=& 0
\end{eqnarray}
This results from the fact that $S(XYZ)=0$ implies $S(XY)=S(Z)$,
$S(XZ)=S(Y)$, and $S(YZ)=S(X)$, as a consequence of the Schmidt decomposition
of the state of $XYZ$.

\end{multicols}
\end{document}